\documentclass[twocolumn]{aastex631}
\usepackage{ifthen}
\usepackage{xspace}
\usepackage{upgreek}
\usepackage{nicefrac}
\usepackage{bm}
\usepackage{ulem}
\usepackage{xspace}
\usepackage{physics}

\usepackage{pgffor}
\makeatletter
\def\dif{\@ifnextchar[{\@with}{\@without}}
\def\@with[#1]#2{
  \ensuremath{\frac{\foreach \x in {#2}{\mathrm{d}\x\,}}{\foreach \x in {#1}{\mathrm{d}\x\,}}}
}
\def\@without#1{
  \ensuremath{%
    \ifx\hfuzz#1\hfuzz
    \mathrm{d}
    \else
    \foreach \x in {#1}{\mathrm{d}\x\,}
    \fi
    }
}
\makeatother
\newcommand{\unit}[1]{\ensuremath{\,\mathrm{#1}}}

\newcommand{\ergs}{\ensuremath{\mathrm{erg\,s^{-1}}}}

\newcommand{\be}{\begin{equation}}
\newcommand{\ee}{\end{equation}}
\newcommand{\ba}{\begin{eqnarray}}
\newcommand{\ea}{\end{eqnarray}}

\newcommand{\ve}{\ensuremath{\varepsilon}}

\newcommand{\G}{\ensuremath{\Gamma_\textsc{k}}}
\newcommand{\E}{\ensuremath{\bm{E}}}
\newcommand{\B}{\ensuremath{\bm{B}}}

\newcommand{\mysub}[1]{\ensuremath{_{\mathrm{#1}}}}

\newcommand{\myerror}[2][NONE]{%
  \ifthenelse { \equal {#1} {NONE} } %
  {\ensuremath{\pm #2}}%
  {\ensuremath{_{-#1}^{#2}}}%  
}

\newcommand{\hess}{H.E.S.S.\xspace}
\newcommand{\magic}{MAGIC\xspace}
\newcommand{\swift}{{\itshape Swift}-XRT\xspace}
\newcommand{\grbHII}{GRB190829A\xspace}

\definecolor{dg}{rgb}{0.0, 0.6, 0.1}
\definecolor{ed}{rgb}{1.0, 0.6, 0.1}
\makeatletter
\def\Andrew{\def\xx{dg}\@ifnextchar[{\@mwith}{\@mwithout}}
\def\Mitya{\def\xx{orange}\@ifnextchar[{\@mwith}{\@mwithout}}
\def\Felix{\def\xx{ed}\@ifnextchar[{\@mwith}{\@mwithout}}
\def\@mwith[#1]#2{\textcolor{\xx}{\sout{#1}#2}}
\def\@mwithout#1{\textcolor{\xx}{#1}}
\makeatother

\newcommand{\dias}{{Dublin Institute for Advanced Studies, School of Cosmic Physics, 31 Fitzwilliam Place, Dublin 2, Ireland}}

\newcommand{\mpik}{{Max-Planck-Institut f\"ur Kernphysik, Saupfercheckweg 1, 69117 Heidelberg, Germany}}
\newcommand{\rikkyo}{{Graduate School of Artificial Intelligence and Science, Rikkyo University, Nishi-Ikebukuro 3-34-1, Toshima-ku, Tokyo 171-8501, Japan}}

\newcommand{\desy}{{DESY, D-15738 Zeuthen, Germany}}

\def\rsize{\ensuremath{10^{16}}}
\def\mygamma{\ensuremath{ 10}}
\def\myL{\ensuremath{10^{39}}}
\def\bone{\ensuremath{ 1}}
\def\btwo{\ensuremath{10^{-3}}}
\def\kappaone{\ensuremath{ 0.90}}
\def\kappatwo{\ensuremath{ 0.10}}
\def\myalpha{\ensuremath{ 2.2}}

\def\myeta{\ensuremath{10^{ 2}}}
\def\myetar{\ensuremath{ 1}}
\shorttitle{Hard VHE emission from SSC sources}
\shortauthors{Khangulyan et al.}
\begin{document}

\title{The formation of hard VHE spectra from GRB afterglow via Two-Zone Synchrotron Self-Compton Emission}

\correspondingauthor{Dmitry Khangulyan}
\email{d.khangulyan@rikkyo.ac.jp}

\author[0000-0002-7576-7869]{Dmitry Khangulyan}
\affiliation{\rikkyo}

\author[0000-0001-9473-4758]{Andrew M. Taylor}
\affiliation{\desy}

\author[0000-0003-1157-3915]{Felix Aharonian}
\affiliation{\dias}
\affiliation{\mpik}



\begin{abstract}
Electron Compton scattering of target photons into the gamma-ray energy band (inverse Compton scattering --IC--) is commonly expected to dominate the very high energy spectra in gamma-ray bursts especially during the afterglow phase. For sufficiently large center-of-mass energies in these collisions, the effect of the electron recoil starts reducing the scattering cross section (the Klein-Nishina regime). 
The IC spectra generated in the Klein-Nishina regime is softer and has a smaller flux level compared to the synchrotron spectra produced by the same electrons.
The detection of afterglow emission from nearby \grbHII in the very high energy (VHE) domain with \hess has revealed an unexpected feature: the slope of the VHE spectrum matches well the slope of the X-ray spectra, despite expectations that for the IC production process, the impact of the Klein-Nishina effect should be strong. The multi-wavelength spectral energy distribution appears to be inconsistent with predictions of one-zone synchrotron-self-Compton models. We study the possible impact of two-zone configuration on the properties of IC emission when the magnetic field strength differs considerably between the two zones. Synchrotron photons from the strong magnetic field zone provide the dominant target for cooling of the electrons in the weak magnetic field zone, which results in a formation of hard electron distribution and consequently of a hard IC emission. We show that the two-zone model can provide a good description of the X-ray XRT and VHE \hess data.
\end{abstract}


\section{Introduction}\label{sec:intro}
The very high energy (VHE; \(>100\unit{GeV}\)) emission detected from gamma-ray burst (GRB) afterglows with \hess and \magic\citep{2019Natur.575..464A,2019Natur.575..455M,2019Natur.575..459M,2021Sci...372.1081H} is considered by many to have inverse Compton (IC) origin \citep[see, e.g,][]{2019Natur.575..448Z}. The emission component produced by
relativistic protons is expected to have a significantly lower flux, due to the very low radiative efficiency of hadronic interactions \citep[see, e.g.,][]{2019Natur.575..464A}. If the VHE emission is produced by relativistic electrons, then because of the so-called synchrotron burn-off limit \citep{1983MNRAS.205..593G} the synchrotron component is expected to reach the VHE regime only if the bulk Lorentz factor is very high, \(\Gamma\geq10^3\). Such high bulk Lorentz factors are excluded during the afterglow phase by energy conservation arguments \citep[e.g., related to self-similar solution for relativistic blast wave obtained by][]{1976PhFl...19.1130B} making IC scattering the most feasible radiation mechanism for the VHE GRB emission during the afterglow period. However, the hard intrinsic spectral slope inferred from observations by \hess of \grbHII afterglow cannot be easily reproduced with standard IC models \citep[see, e.g.,][]{2021Sci...372.1081H}. This leaves one of two possibilities: (i) invoke alternative radiation mechanisms, or (ii) develop a more sophisticated IC scenario to provide a better description of the observational data.

Synchrotron radiation is a very efficient radiative emission mechanism of electrons during the afterglow phase of GRBs. If the synchrotron component extends into the VHE domain, it can reproduce the flux level and spectral slope revealed with \hess from \grbHII afterglow \citep{2021Sci...372.1081H}. While the conservation of energy, used to constrain the bulk Lorentz factor, is a robust argument, the burn-off energy limit can be avoided in certain non-standard scenarios. For example, if the strength of the accelerating electric field, \({\cal E}\), exceeds the strength of the magnetic field, \(B\) (in a plasma such configurations require non-ideal magnetohydrodynamics) then synchrotron emission can extend beyond the burn-off limit by the factor of \({\cal E}/B\).
Alternatively, in highly turbulent magnetic fields magnetobremsstrahlung radiation can extend beyond the burn-off limit \citep{2013ApJ...774...61K}. 
{If the correlation length of the magnetic field is large compared to the photon formation length, \(m_e c^2/e\bar{B}\) (here \(m_e\) and \(e\) are electron mass and charge, respectively; \(c\) and \(\bar{B}\) are light speed and averaged magnetic field), then the radiation is generated in the synchrotron regime, resulting in the burn-off limit for the synchrotron maximum energy \citep[for a detailed consideration, see, e.g., in][]{2013ApJ...774...61K,2019ApJ...887..181D}. However, if the correlation length is short compared to the photon formation length, then the electrons instead emit in the jitter regime, and the emission peaks at higher energy compared to the synchrotron case, alleviating the limit from the burn-off limit \citep{2013ApJ...774...61K}.}
Finally, the electron synchrotron spectrum can extend beyond the burn-off limit in two-zone systems, where the physical conditions at the acceleration site and in the radiation production region differ substantially
\citep{2012MNRAS.427L..40K,2021ApJ...914...76K}.
In conclusion, there are several ways of expanding the energy spectrum of magneto-bremsstrahlung to high or even very high energies. However, the feasibility of these scenarios depends on the implementation of many factors and requires extreme assumptions. 

In contrast, IC scattering is a natural and very effective channel of VHE gamma-ray production. Although the recent observations of VHE gamma rays during the GRB afterglows challenge the simple one-zone IC model, more sophisticated scenarios cannot be excluded. In this paper, we study the spectral properties of gamma rays in the two-zone IC model in which the production region of the target (synchrotron) photons and the IC gamma-ray emitter are separated. One can propose several possible realizations for such a two-zone setup. For example, one may expect quite different conditions at the forward and reverse shocks, which propagate through the circumburst medium (CBM) and the jet, respectively. If the emission from the reverse shock appears to be important at certain frequencies then a two-zone description for GRB afterglow emission should be considered \cite[see, e.g.,][]{2022MNRAS.512.2337D,2022ApJ...931L..19S}. Alternatively, the shock region itself can be quite complex potentially providing quite different physical conditions for particle acceleration and radiation. Indeed, simulations suggest that downstream shock material, the dominant emission site during the afterglow phase, is expected to be highly inhomogeneous, an aspect usually neglected in GRB afterglow emission modelling. Below we consider the impact of a strongly inhomogeneous magnetic field on the properties of IC emission. We show that under reasonable assumptions, even a two-zone synchrotron self-Compton (SSC) scenario can provide a considerably improved description of the broadband spectra reported from \grbHII.

\section{Standard One-Zone SSC Scenario}\label{sec:grb}

The standard GRB
afterglow emission framework postulates that this emission is generated via the synchrotron and IC channels, with
synchrotron radiation providing the dominant target for IC scattering -- the so called SSC
scenario. The analysis of the spectral energy distribution (SED) in SSC models is straightforward if the IC emission is
generated in the Thomson regime \citep[see, e.g.][]{2001ApJ...548..787S}, as in this case the energy loss rate,
\(\dot{E}\), has a simple form \(\dot{E}\propto E^2\) (here \(E\) is electron energy). In this regime, a power-law
injection of non-thermal electrons, \(q\propto E^{-\alpha}\) (here \(\alpha\) is the injection index, for conventional acceleration mechanisms one typically assumes \(\alpha\approx2\)), leads to the formation of a broken-power-law distribution of radiating electrons. The
synchrotron and IC (Thomson) components generated by these electrons also reflect this broken-power-law shape,
with the IC component dominating at higher energies. The subsequent broadband SED produced is double-humped, with the relative emissivity of the synchrotron and IC components being determined by phenomenological parameters (typically, by the radiation efficiency, i.e., by the fraction of energy radiated away). The photon index of the synchrotron spectrum, produced by electrons with energies above the cooling break, is \(\gamma\mysub{s}=(\alpha+2)/2\), provided that \(\alpha>1\).
In the single zone SSC scenario, the corresponding IC spectrum has the same photon index, if generated in the Thomson regime.

\begin{figure}
  \plotone{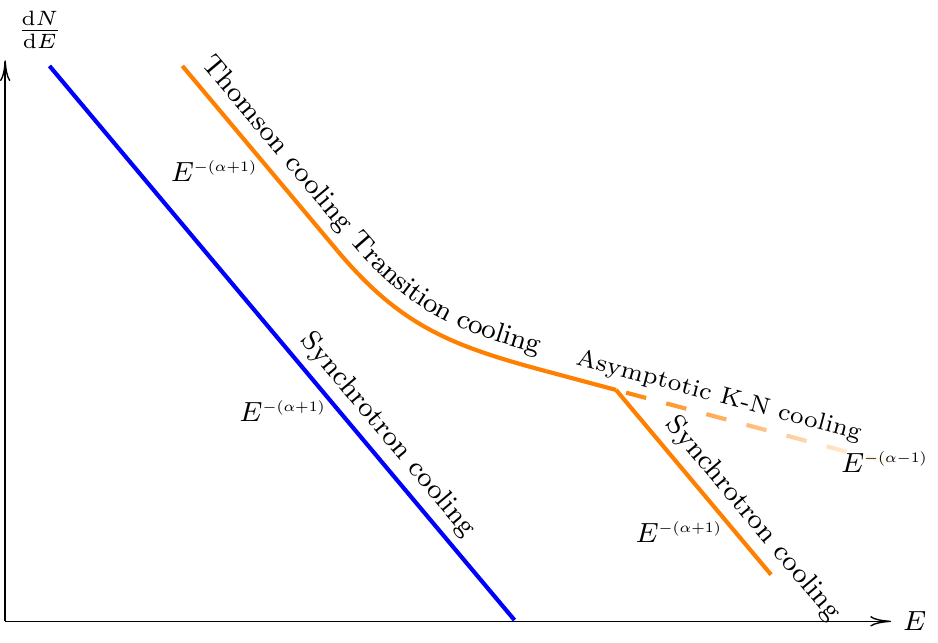}
  \caption{A sketch that illustrates the formation of the particle spectrum in the case of dominant synchrotron losses and dominant IC losses. The part of the spectrum formed in the fast cooling regime is shown. \label{fig:sketch1}}
\end{figure}

\begin{figure}
  \plotone{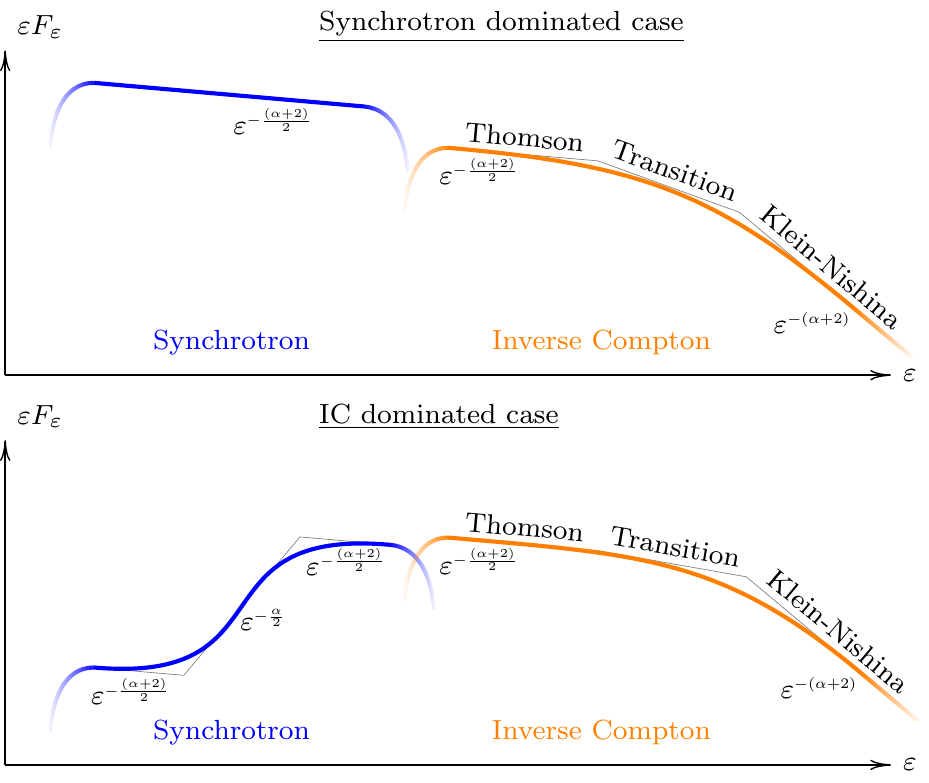}
  \caption{A sketch that illustrates the formation of the SED in the case of dominant synchrotron losses and dominant IC losses.  \label{fig:sketch2}}
\end{figure}

Typically, during the afterglow phase the (synchrotron) X-ray spectrum is observed to be hard, with a photon index \(\sim2\). Thus the photons detected in the X-ray band provide a non-negligible target for IC scattering. In the plasma co-moving frame, the energy of the electron, $E$, generating the VHE emission, detected at energy\footnote{Note that we prime the quantities in the progenitor frame, and we neglect the cosmological redshift effect.} \(\ve\mysub{vhe}'\), satisfy the condition: \(E>\ve\mysub{vhe}'/\Gamma\).
If electrons of this energy up-scatter photons from a component detected by the observer at energy
\(\ve\mysub{x}'\), then the typical product of the target photon and electron energies, which determines the scattering regime, is
\be
\frac{E\ve\mysub{x}}{m_e^2c^4}> \frac{\ve\mysub{vhe}'\ve\mysub{x}'}{m_e^2c^4\Gamma^2}\approx4
\qty(\frac{\Gamma}{10})^{-2}\left(\frac{\ve\mysub{vhe}'}{0.1\unit{TeV}}\right) \left(\frac{\ve\mysub{x}'}{1\unit{keV}}\right)\,.
\ee
Here \(m_e\) and \(c\) are the electron mass and speed of light, respectively. Unless the bulk Lorentz factor is high, \(\Gamma\geq10^2\), the electrons that produce the VHE emission up-scatter a considerable part of the photon targets in the Klein-Nishina regime. The study of the VHE properties of GRB afterglows should therefore be conducted with models that account for the change of the IC cross-section in the relativistic regime.

The influence of the Klein-Nishina regime on the SED is two-fold, as one must account for both the change of the emission and energy loss rates \citep[see, e.g.,][]{2003AIPC..662..292D,2009ApJ...703..675N}. In the fast cooling regime, the particle spectrum, \(\dd{N}=n\dd{E}\), is determined by the injection spectrum, \(q\), and by the cooling time \(\tau=E/|\dot{E}|\):
\be
n(E) = \frac{\tau (E)}{E}\int\limits_{E}^{\infty} \dd{\tilde{E}}\,q\qty(\tilde{E})\,.
\ee
If the injection is a power-law \(q\propto E^{-\alpha}\), then the particle spectrum is 
\be\label{eq:particles_simple}
n(E) \propto \tau(E) E^{-\alpha}\,.
\ee
(Note that here we assume that the injection spectrum is sufficiently steep so as to ensure the integral is dominated by the low energy limit).

If the synchrotron losses dominate over the Compton losses (more specifically if the energy density of the magnetic field is larger than the energy density of the target photons) then $\tau(E) \propto E^{-1}$, and a power-law spectral injection also yields a power-law distribution of particles: \(n(E)\propto E^{-(\alpha+1)}\) (see Fig.~\ref{fig:sketch1} for a sketch of the cooled particle spectrum). Subsequently, a power-law synchrotron component is produced with  photon index \(\gamma\mysub{s}\).

The inverse Compton of radiation has the same power-law photon index as long as the scattering takes place in the Thompson regime.
In the Klein-Nishina regime, the IC slope should (asymptotically, i.e., ignoring the logarithmic term) approach \(\gamma\mysub{kn}\approx(\alpha+2)\) \citep[provided that the emitting electrons obey a power-law energy distribution with index \(\alpha+1\),][]{1970RvMP...42..237B}. Thus, since the slope of IC component generated in the Thomson regime matches that of the synchrotron radiation, \(\gamma\mysub{s}\), the Klein-Nishina effect causes a spectral softening by \(\Delta\gamma\approx\gamma\mysub{kn}-\gamma\mysub{s}\approx(\alpha+2)/2\). For example, if \(\alpha\approx2\) then the spectral slope changes from \(\gamma\mysub{s}\approx2\) to \(\gamma\mysub{kn}\approx4\), and the spectral softening is \(\Delta\gamma\approx2\). A schematic of the SED is shown in Fig.~\ref{fig:sketch2}. One should note that for a broad target photon distribution, the transition to the Klein-Nishina regime is spread over a  broad energy range and can have a rather complex character.

The situation changes dramatically when the energy density of target photons is larger than the energy density of the magnetic field. In this case, the impact of the Klein-Nishina effect on the formation of the  electron spectrum becomes a dominant factor. The radiative cooling time \(\tau(E)\) can be approximated by a broken power-law function: for sufficiently low electron energies, the IC interaction proceeds in the Thomson regime, thus 
$\tau(E)\propto E^{-1}$. At higher energies, the IC interactions occur in the Klein-Nishina regime where the energy loss rate is energy-independent, thus
$\tau(E)\propto E$. Finally at even higher energies, denoted \(E_*\), the synchrotron losses (as their rate increases with particle energy) begin to dominate over the IC energy losses, and the original energy dependence of the cooling time is recovered:
$\tau(E)\propto E^{-1}$. As follows from Eq.~(\ref{eq:particles_simple}), for a power-law injection spectrum, the particle spectrum formed in the fast cooling regime should also be a double-broken-power-law (with the power-law index changing as \(\alpha+1\rightarrow \alpha-1\rightarrow \alpha+1\): see  Fig.~\ref{fig:sketch1}). The \(E^{-(\alpha+1)}\) part of the spectrum formed under dominant (Thomson regime) IC losses changes to, \(\propto E^{-(\alpha-1)}\), formed under the dominant IC (Klein-Nishina regime) losses. Finally, above \(E_*\), the spectrum softens back to \(E^{-(\alpha+1)}\). We note, however, that the transition to the Klein-Nishina regime proceeds smoothly, therefore the spectrum does not follow precisely the schematic shape explained above. For example, as can be seen from Fig.~\ref{fig:cooling}, the IC cooling time in the transition regime is better approximated as a constant, \(\tau\approx\mathrm{const}\). Therefore, the corresponding transformation of the electron spectrum is better approximated as \(\alpha+1\rightarrow \alpha\rightarrow \alpha+1\) (note that this power law index is indicated in the bottom panel of Fig.~\ref{fig:cooling} with a black guide line). 

As for the  synchrotron radiation, electrons cooled by IC in the Thomson regime produce a spectrum with photon index \(\gamma\mysub{s}\); at higher energies the hardening of the electron spectrum due to the dominant Klein-Nishina energy losses results in a hard synchrotron spectrum with photon index in the range between \(\gamma\mysub{s}\) and \(\gamma\mysub{s,kn}\approx\alpha/2\) (\(\gamma\mysub{s,kn}\) is the limiting value achieved under IC cooling in the deep Klein-Nishina regime: see Fig.~\ref{fig:sketch2}). In the transition region with an approximately constant IC cooling time, the slope of the synchrotron spectrum is approximately  \((\alpha+1)/2\), as indicated by the black guide lines in Figs.~\ref{fig:SED}~and~\ref{fig:SED_ab}. Finally, the emission produced by electrons with energies exceeding \(E_*\) has the standard synchrotron slope \(\gamma\mysub{s}\). As the synchrotron and IC energy loss rates for particles with \(E_*\) are equal, the narrow-band luminosity of the synchrotron and IC components produced by particles with \(E_*\) are (almost) equal.

The spectral shape of the IC component is different to that of the synchrotron spectrum. The component generated in the Thomson regime has a spectral index of \(\gamma\mysub{s}\). At higher energies the impact of the Klein-Nishina effect on the particle spectrum is partially compensated by the reduction of the cross section. For example, in the limiting regime, a spectrum \( \propto E^{-(\alpha-1)}\)  generates in the Klein-Nishina regime a \(E^{-\alpha}\) IC spectrum. For \(\alpha\approx2\) a  Thomson spectrum with photon index \((\alpha+2)/2\) transits smoothly into the Klein-Nishina spectrum with photon index \(\alpha\). However, in the region of transition to the Klein-Nishina regime, this asymptotic photon index might be quite a coarse approximation. Moreover,  above \(E_*\)  the synchrotron losses dominate, thus the Klein-Nishina spectrum eventually softens to \(\alpha+2\) above \(E_*\). Note that in the Klein-Nishina regime almost all the electron energy is transferred to the up-scattered photon, so the photon energy in the co-moving frame is equal to that of the incident electron energy, \(\ve\mysub{ic}\approx E_*\).

Observations of \grbHII with \hess revealed that VHE component, corrected for the extragalactic background light (EBL) attenuation, is best described as a single power-law spectrum extending up to \(3\unit{TeV}\) with a hard photon index of \(\gamma\mysub{vhe}=2.07\myerror{0.09}\) \citep{2021Sci...372.1081H}. Strikingly, this slope matches well the slope of the X-ray spectrum measured with \swift \citep[e.g., \(\gamma\mysub{xrt}=2.03\myerror{0.06}\) during the first night][]{2021Sci...372.1081H}.
Also, the \swift and \hess observations revealed that the fluxes in the X-ray and VHE bands appeared to be similar \citep[potentially a natural feature of pair loading feedback, see][for detail]{2016MNRAS.460.2036D,2019ApJ...880L..27D}.

In the VHE band the influence of the Klein-Nishina effect should be noticeable. However, this spectral effect was not observed in the \hess measurements. In the framework of the simple one-zone analysis introduced above, the slope and flux level match implies that the cooling of TeV emitting electrons proceeds in the Klein-Nishina regime, and that the X-ray synchrotron is produced by particles with energy exceeding \(E_*\). As the hard VHE spectrum extends up to \(3\unit{TeV}\), then \(E_*>0.3\qty(\Gamma/10)^{-1}\unit{TeV}\). The synchrotron emission produced by the high-energy electrons is detected by the observer at
\be
\ve\mysub{syn}'>60\qty(\frac{\Gamma}{10})^{-1}\frac{B}{1\unit{G}}\unit{keV}\,.
\ee
This estimate shows that a very low magnetic field of \(\sim\unit{mG}\) level is required by the VHE measurements. Such a low magnetic field, however, is incompatible with the required radiation efficiency of the production region given the adiabatic cooling time is \(\tau\mysub{ad}\sim t\mysub{tr}'\Gamma\), where \(t\mysub{tr}'\) is time since the GRB trigger (as measured by a distant observer at rest in the progenitor reference frame). The broad-band SED obtained with \swift and \hess therefore cannot be reproduced in the framework of the standard one-zone SSC scenario \citep[see also][]{2022ApJ...925..182H}. To resolve the spectral issue in SSC scenario one needs either: (1) assume that there is an important low-energy target photon field, probably of external origin; or (2) consider a two-zone scenario.

The former scenario requires the presence of an external target that provides a target of an energy density comparable to that of the magnetic field in the plasma co-moving frame:
\be
w\mysub{ext}\sim 4\times10^{-2}\qty(\frac{B}{1\unit{G}})^2\unit{erg\,cm^{-3}}\,.
\ee
If the photons are isotropic in the progenitor frame, then we obtain \(w\mysub{ext}'\sim 4\times10^{-4}\qty({10B}/{(\Gamma\unit{G}}))^2\unit{erg\,cm^{-3}}\). The VHE emission detected from \grbHII lasted for almost \(\Delta t=50\)~h \citep{2021Sci...372.1081H}, and the forward shock covered a distance of \(\Delta R'\sim \Gamma^2\Delta t c\sim 10^{17}\qty(\Gamma/10)^2\unit{cm}\). The luminosity of the photon field should therefore be
\be
L\mysub{ext}'\sim 4\pi \Delta R'{}^2 w\mysub{ext}'c\sim 10^{42}\qty(\frac{B}{1\unit{G}})^2\unit{erg\,s^{-1}}\,.
\ee
If the magnetic field is weak, \(B\ll1\unit{G}\), then an external photon field of reasonable luminosity can provide a sufficiently dense external photon field \citep[see, e.g.,][]{2021ApJ...920...55Z}, however external IC scenarios with an equivalent Gauss-strength magnetic field cannot be realized.

\section{Two-Zone SSC Emission Scenario}\label{sec:scenario}

\subsection{Physical justification}\label{sec:setup}
\begin{figure}
  \plotone{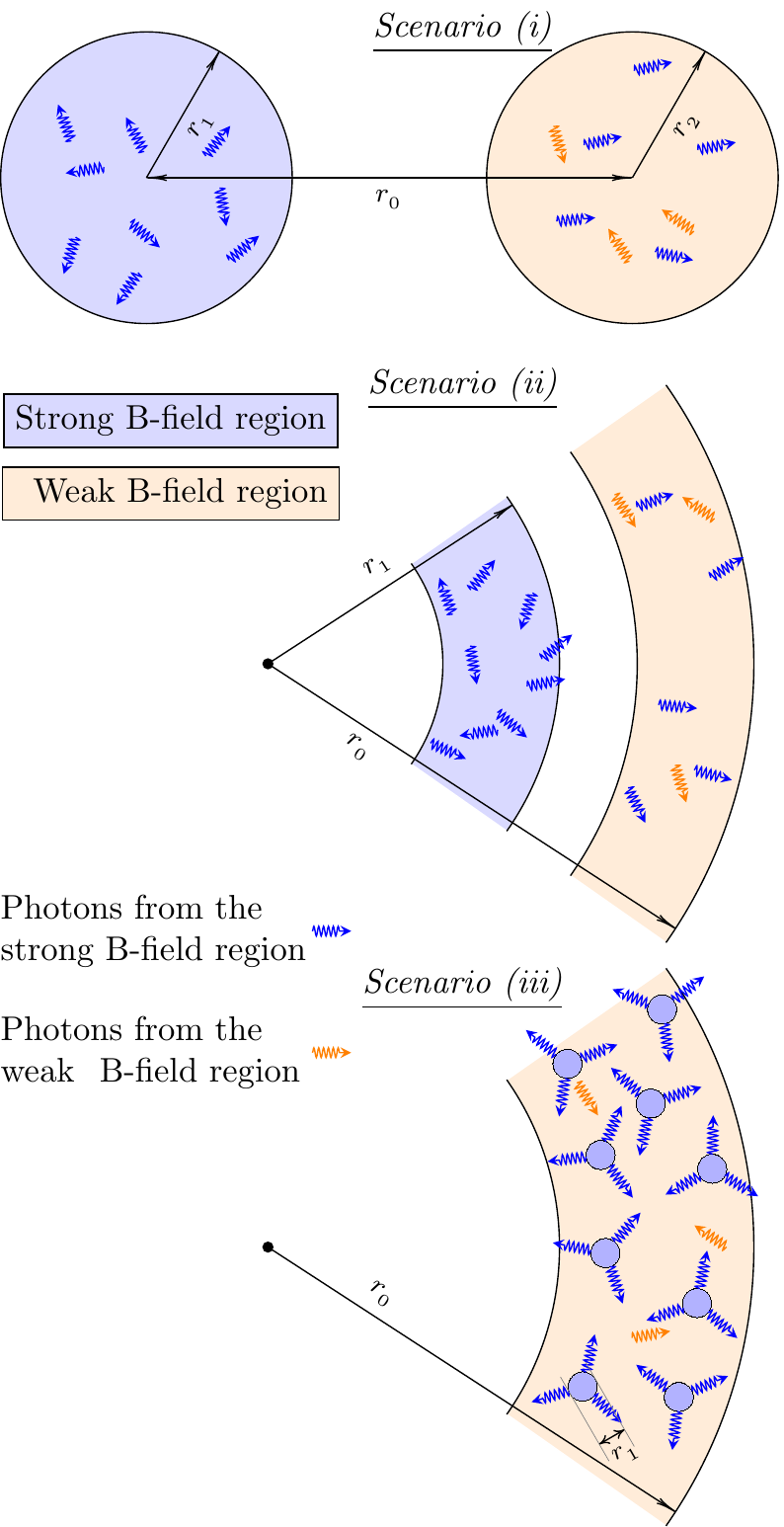}
  \caption{Examples of three different geometries that allow the scenario realization. Scenario (i):  two distinct regions with typical sizes of \(r_1\) and \(r_2\) separated by a distance \(r_0\); scenario (ii): two converging shells of radius \(r_1\) and \(r_0\); scenario (iii): a large number of compact regions (of typical size  \(r_1\)) with strong magnetic field embedded within a larger zone of size \(r_0\).  \label{fig:three}}
\end{figure}

We consider the emission region consisting of two zones: the first zone with a strong magnetic field, $B_{1}$, and the second zone with a weak magnetic field, $B_{2}$, with \(B_1\gg B_2\).  Should particles themselves also easily mix between the two zones, then one would not expect a significant difference between the energy distributions of particles in these zones. We here, however, assume that the particle exchange between the zones is inefficient, and thus two distinct particle distributions, \(n_1\) and \(n_2\), are formed in the two zones.

The target photons, however, travel freely between the two zones. The specific realization of the scenario, in particular the shapes and relative location of the zones, determines the actual distribution of target photons in the zones. Let us qualitatively consider several possible realizations of the two-zone scenario: (i) two distinct regions with typical sizes of \(r_1\) and \(r_2\) separated by a distance \(r_0\); (ii) two converging shells of radius \(r_1\) and \(r_0\); (iii) \(N\) compact regions (of typical size  \(r_1\)) with strong magnetic field embedded within a larger zone of size \(r_0\). These three possibilities are shown in Fig.~\ref{fig:three}. Although less apparent, scenario (iii) is two-zone in the sense that the physical conditions and processes are the same in the compact regions, and differ substantially from those in the larger zone.

The synchrotron luminosity of each of the zones is \(L_1\) and \(L_2\), respectively. In scenario (iii) we define \(L_1\) as the total luminosity of \(N\) regions of enhanced  B-field. We consider a situation \(L_1\gg L_2\). Thus, when considering the processes in the first zone, we can ignore the photons supplied by the second zone. The energy density of the locally generated photons in the first zone is
\be \label{eq:target_density}
w_{1\rightarrow1} \sim \frac{L_1}{r_1^2 N c}\,,
\ee
where \(N=1\) for scenarios (i) and (ii). Equation~\eqref{eq:target_density} ignores a numerical factor, which depends on the production region geometry and the distribution of emitting particles. For example, in the case of a spherical homogeneous production region, the volume average energy density of target photons is given by  Eq.~\ref{eq:target_density} with a factor \(9/(16\pi)\) \citep[for detail, see in][]{1996MNRAS.278..525A}. We note that such factors do not affect our conclusions, we therefore safely ignore them.

In the second zone one needs to account for the contribution of locally generated photons:
\be
w_{2\rightarrow2}^{(i)} \sim \frac{L_2}{r_2^2 c}\qq{and}w_{2\rightarrow2}^{(ii)/(iii)} \sim \frac{L_2}{r_0^2 c}
\ee
and the photons supplied from the first zone, \(w_{1\rightarrow2}\). For the each of the above defined geometries one obtains
\be
w_{1\rightarrow2} \sim \frac{L_1}{r_0^2 c}\,.
\ee
The suggested scenario is realized if the photon field produced in the first zone (being locally a subdominant)  provides the dominant target for the particle cooling in the second zone:
\be
\label{eq:conditions}
w_{1\rightarrow1}\ll\frac{B_1^2}{8\pi}\qq{and} w_{1\rightarrow2}\gg\frac{B_2^2}{8\pi}\,.%,\quad w_{1\rightarrow2}\gg w_{2\rightarrow2},
\ee
The photon field in the second zone is diluted compared to the first zone: \(w_{1\rightarrow1}>w_{1\rightarrow2}\), thus
the scenario requires that \(B_1\gg B_2\). The difference of the magnetic fields determines the dilution of the
photon field, \(\kappa = w_{1\rightarrow2}/w_{1\rightarrow1}\), that allows the scenario realization (i.e, the conditions given by Eq.~\ref{eq:conditions}).

The possible ratio of the magnetic fields should be determined by the physical arguments unique to each specific realization of the
scenario. However, from the general point of view, it is obvious that if the photon field is significantly diluted in
the second zone, \(\kappa\ll1\), the required difference between the magnetic field strength becomes larger, making the
realization of the scenario less feasible (although not excluded). For example, a strong dilution might be expected in scenario (i)
provided that \(r_0\gg r_1\). In contrast, in scenario (ii) the dilution of the photon field in the second zone is
small, by a factor of \(\sim2\), provided that two shells are of comparable radius, \(r_1\approx r_0\). Similarly, in
scenario (iii) one obtains
\be
\frac{w_{1\rightarrow2}}{w_{1\rightarrow1}}\sim \frac{r_1^2 N}{r_0^2}=\frac{fr_0}{r_1}\,,
\ee
where \(f\) is filling factor. If the above ratio is not small (i.e., \({w_{1\rightarrow2}}/{w_{1\rightarrow1}}\gtrsim1\)) then the photon field is nearly homogeneous in the entire production region, i.e., \(\kappa\approx 1\).

For the sake of simplicity we will consider a single common photon target being present in the two zones. In the first place, this seems to be a perfectly suitable choice for scenarios (ii)  and (iii)  if  \(r_1\approx r_0\) and \(fr_0/r_1\gtrsim1\), respectively. Even if these conditions are not fulfilled, the model calculations should reproduce  correctly the part of SED formed in the fast cooling regime (provided that IC losses dominate over the synchrotron cooling in the second zone: \(w_{1\rightarrow2}\gg B_2^2/(8\pi)\)).

Although the scenario can be realized also in scenario (i), if the magnetic field in the second zone is sufficiently weak to remain subdominant compared to the significantly diluted photon field provided from the first zone,  scenarios (ii) and (iii) seem to be less demanding. In particular, these geometries can be formed during the afterglow phase of GRBs. The shells assumed in scenario (ii) may correspond to the reverse and forward shocks. Also an onion-like structure may be formed in the inner part of the forward shock downstream region, where the competing processes of magnetic field amplification and decay may lead to the formation of a layer with an enhanced magnetic field. If the magnetic field amplification in the downstream proceeds in a highly non-homogeneous manner, then instead of a shell-like structure one should expect rather a large number of magnetized blobs in the production region, i.e., scenario (iii). Although scenarios (ii) and (iii) are characterized by quite similar geometries, the angular distribution of the target photons in the second zone may be quite different in these two cases. While in scenario (iii), the target photons are nearly isotropic,  scenario (ii) features a substantial anisotropy of the target photons in the second zone (as depicted in Fig.~\ref{fig:three}). As the emitting particles are isotropized in the plasma frame, this photon anisotropy should not have any impact on the cooling process. However, one may need to account for anisotropic IC cross-section \citep[see, e.g.,][]{1981Ap&SS..79..321A} for accurate computation of the IC spectra. For example, if the emission generated in the direction of the observer is predominately produced by scattering target photons at small scattering angles, then the IC spectra appear to be harder compared to the spectra computed with angle-averaged IC cross-section \citep[see, e.g.,][]{2008MNRAS.383..467K}.

Because of the Doppler boosting effect, the observer can detect the emission coming from a patch of the shell with a typical size of  \(R'/\Gamma\), where \(R'\sim t'\mysub{tr}\Gamma^2c\). Thus, one obtains the patch size as \(t'\mysub{tr}\Gamma c\gg10^{15}\unit{cm}\) (provided that \(t'\mysub{tr}>1\unit{h}\) for the afterglow period). The realization of scenario (iii) requires that the size of the blobs is small, \(r_1\ll t'\mysub{tr}\Gamma c\). Verification of this condition from the first principles may require detailed plasma simulations, which are beyond the scope of this study. As in the case of GRB afterglow, the GeV emission seems to belong to the same component as the synchrotron, we may therefore speculate that the acceleration in the blobs are limited by the synchrotron cooling and the acceleration process is efficient, \(\eta\mysub{acc}\sim1\) (here \(\eta\mysub{acc}\) is the acceleration efficiency). Thus, the size of the blobs should be sufficiently large to confine particles with energy
\be
E\approx 60\left(\frac{B}{1\unit{G}}\right)^{\nicefrac{-1}{2}}\unit{TeV}\,.
\ee
The corresponding gyro radius,
\be
R_{\cal{G}}\approx 2\times 10^{11} \left(\frac{B}{1\unit{G}}\right)^{\nicefrac{-3}{2}} \unit{cm}\,,
\ee
is significantly smaller than the patch size. This likely implies that there are no fundamental constrains from the plasma physics forbidding the scenario realization.

\subsection{Mathematical setup}\label{sec:math}

For each of the zones we consider the injection-cooling equation:
\be
\pdv{n_1}{t} + \pdv{\dot{E}_1n_1}{E} = q_1(E) - \frac{n_1}{\tau_{1\rightarrow2}}+\frac{n_2}{\tau_{2\rightarrow1}}\,,
\ee
\be
\pdv{n_2}{t} + \pdv{\dot{E}_2n_2}{E} = q_2(E) + \frac{n_1}{\tau_{1\rightarrow2}}-\frac{n_2}{\tau_{2\rightarrow1}}\,.
\ee
Here \(q_i\) is the injection term; \(\dot{E}_i\) is the energy loss rate in each zone; and \(\tau_{i\rightarrow j}\) is the probability of particle escape from zone \(i\) to zone \(j\). To illustrate the possible impact of the two-zone setup on the IC spectrum we assume that in the energy range of interest \(\tau_{i\rightarrow j}\gg \tau_{1}, \tau_{2}\), where \(\tau_{i}=E/|\dot{E}_i|\) is radiative cooling time in zone \(i\).  Since we are interested in the high-energy part of the spectrum, which is formed in the fast cooling regime, we consider the following simplified equations:
\be
\pdv{\dot{E}_in_i}{E} = q_i(E)\,.
\ee

The magnetic field strengths differ significantly in each zone, thus we do not adopt a universal cutoff energy in the injection spectrum, but instead find different injection rates, $A_{i}$, and cutoff energies, $E_{{\rm cut},i}$, within the two zones. We, however, assume that within both zones the injection function has a common power-law spectral index:%, \(\alpha\sim 2\):
\be\label{eq:injection_spectrum}
q_i=A_i E^{-\alpha} \exp[-\frac{E}{E\mysub{cut}{}_{,i}}]\,,
\ee
where the cutoff energy is found through the balance of acceleration and loss time-scales:
\be\label{eq:max_energy}
\frac{\eta\mysub{acc} E\mysub{cut}{}_{,i}}{eB_ic}=-\frac{E\mysub{cut}{}_{,i}}{\dot{E}_i(E\mysub{cut}{}_{,i}
)}\,.
\ee
Here \(e\) is electron charge. Whilst the acceleration parameter \(\eta\mysub{acc}\) is assumed to be the same in both zones, the energy losses are computed independently for each zone.

We assume that dimensionless parameters \(\kappa_1\) and \(\kappa_2\) (\(\kappa_1+\kappa_2=1\)) define the fraction of the total energy injected into zone one and two, respectively: 
\be\label{eq:normalization}
\frac{1}{\kappa_i} \int\limits_{E\mysub{min}}^{\infty} \dd{E} E q_{i}(E) = L_0\,,
\ee
where \(L_0\) is the total power injected in the production region. The minimum energy, \(E\mysub{min}\), we set to a value of \(\Gamma m_ec^2\) (here \(m_e\) is electron mass and \(c\) is the speed of light). While Eq.~(\ref{eq:max_energy}) defines the cutoff energy, Eq.~(\ref{eq:normalization}) determines the normalization coefficients in each zone, \(A_i\).

As in the high-energy regime, the synchrotron and IC losses are expected to provide the dominant energy loss channels, we therefore only take account of these two energy loss mechanisms. Since the photon field is common between the zones, the difference between the energy loss rate in the zones is due to the different synchrotron losses within each zone:
\be\label{eq:losses}
\dot{E}_i=\dot{E}\mysub{syn}{}_{,i}+\dot{E}\mysub{ic}\,.
\ee
The synchrotron energy losses in zone \(i\) are determined by the magnetic field strength
\be\label{eq:syn_energy_losses}
\dot{E}\mysub{syn}{}_{,i}=-\frac{16\pi}{3}\frac{e^4E^2}{m_e^4c^7}\qty(\frac{2}{3} \frac{B_{i}^2}{8\pi})\,.
\ee
Note that the equation above is averaged over pitch angle. The synchrotron cooling time is 
\be
\tau\mysub{syn}{}_{,i}=\frac{E}{|\dot{E}\mysub{syn}{}_{,i}|}\approx 400\qty(\frac{B_{i}}{1\unit{G}})^{-2}\qty(\frac{E}{1\unit{TeV}})^{-1}\unit{s}\,.
\ee

The IC losses are determined by the energy distribution and number density of target (synchrotron)
photons. As photons can freely cross the zone boundaries we assume that the photon distribution is the same throughout the entire production region, i.e., it includes the contributions from both zones. We compute the synchrotron emission using the particle distribution in each zone and the corresponding magnetic field:
\be
\dv{N\mysub{syn}{}_{,i}}{\ve\dd{t}}\approx\int\limits_{m_ec^2}^{\infty}\dd{E} n_{i} K\mysub{syn}{}_{,\ve}\qty(E,B_{i})\,,
\ee
where \(\ve\) is the target (synchrotron) photon energy. For the synchrotron integral kernel, \(K\mysub{syn}{}_{,\ve}\), we use a simple analytic approximation for the pitch angle averaged synchrotron spectrum \citep[for detail see in][]{2010PhRvD..82d3002A}. Finally, we compute the energy distribution of the target photons as 
\be
\dv{N\mysub{syn}}{\ve\dd{V}}\approx\frac{1}{R^2c}\qty(\dv{N\mysub{syn}{}_{,1}}{\ve\dd{t}}+\dv{N\mysub{syn}{}_{,2}}{\ve\dd{t}})\,.
\ee
Here \(R\) is size of the production region.

The rate of IC scattering is determined by the angle averaged scattering cross section \citep[for detail see in][]{1968PhRv..167.1159J}:
\[
  \dv{\nu\mysub{ic}}{\ve_\gamma\dd{\ve}}=\frac{8\pi cr_0^2}{bE}\dv{N\mysub{syn}}{V\dd{\ve}}\times
\]
\[
\left[1+\frac{z^2}{2(1-z)}+\frac{z}{b(1-z)}-\frac{2z^2}{b^2(1-z)^2}-\right.
\]
\be
\left.\frac{z^3}{2b(1-z)^2}-\frac{2z}{b(1-z)}\,\ln\frac{b(1-z)}{z}\right]\,.
\ee
Here \(r_0=e^2/m_ec^2\) is the electron  classical radius; the  Klein-Nishina parameter is given by $b=4\ve E/(m_e^2c^4)$; and \(z\) is the ratio of the up-scattered photon to electron energy, \(z=\ve_\gamma/E\). The IC energy loss rate depends on the energy distribution of target photons as
\be
\dot{E}\mysub{ic}\approx       \int\limits_0^\infty\dd{\ve}\int\limits_{\ve\mysub{min}{}_{,\gamma}}^{\ve\mysub{max}{}_{,\gamma}}\dd{\ve_\gamma}(\ve-\ve_\gamma)\dv{\nu\mysub{ic}}{\ve_\gamma\dd{\ve}}\,,
\ee
where the maximum/minimum energy of up-scattered gamma-ray, \(\ve\mysub{max/min}{}_{,\gamma}\), is determined by kinematic constraints. If electrons up-scatter low-energy target photons (i.e., the Klein-Nishina parameter is small, \(b\ll1\)), then the IC energy loss rate depends only on the energy density of the target photons, \(w\mysub{ph}\):
\be\label{eq:thomson_energy_losses}
\dot{E}\mysub{T}{}_{,i}=-\frac{32\pi}{9}\frac{e^4E^2}{m_e^4c^7}w\mysub{ph}\,,
\ee
analogous to the corresponding angle averaged energy loss rate in a magnetic field given in Eq.~(\ref{eq:syn_energy_losses}).

\subsection{Model calculations}\label{sec:calcs}

\begin{figure}[ht!]
\plotone{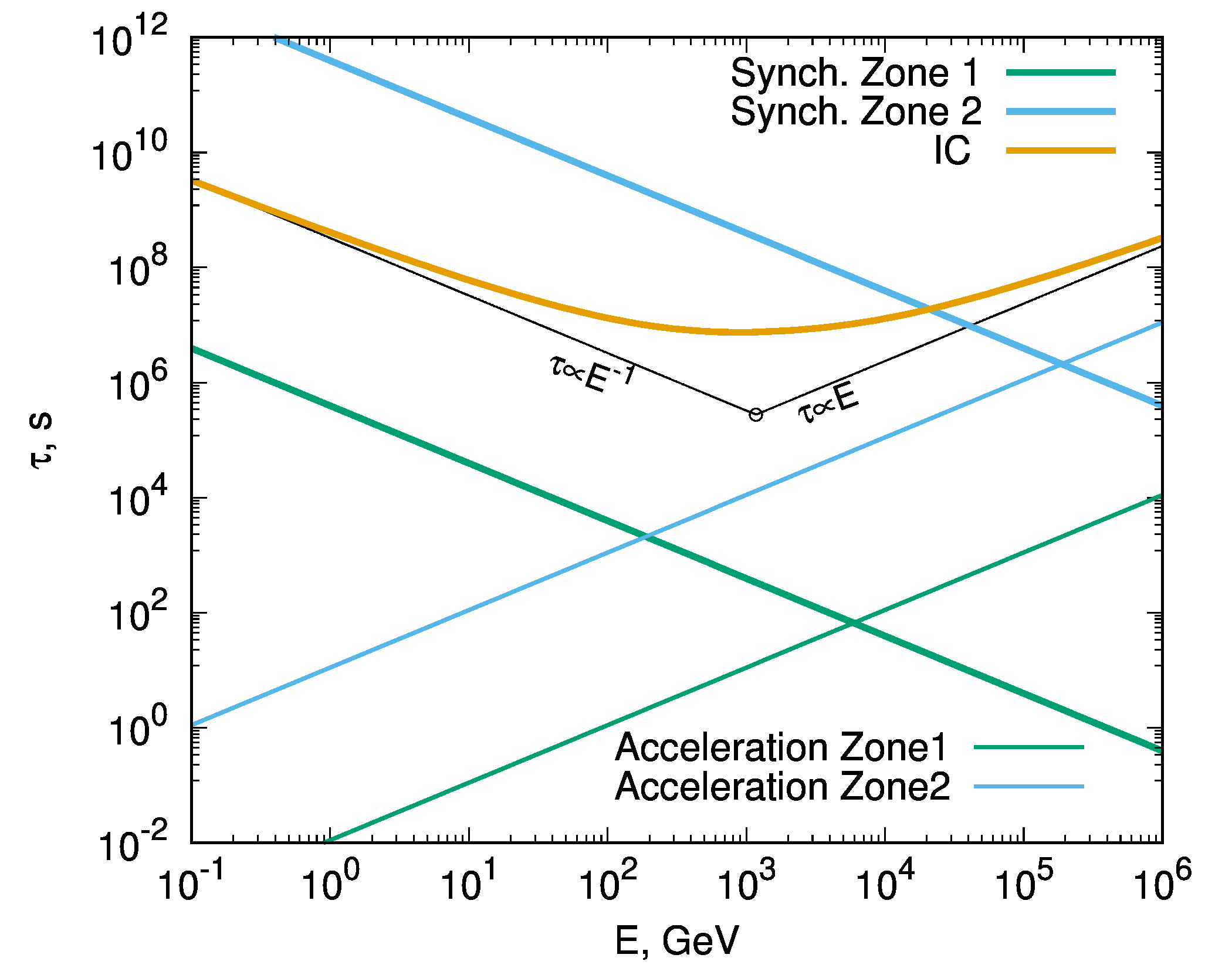}
\plotone{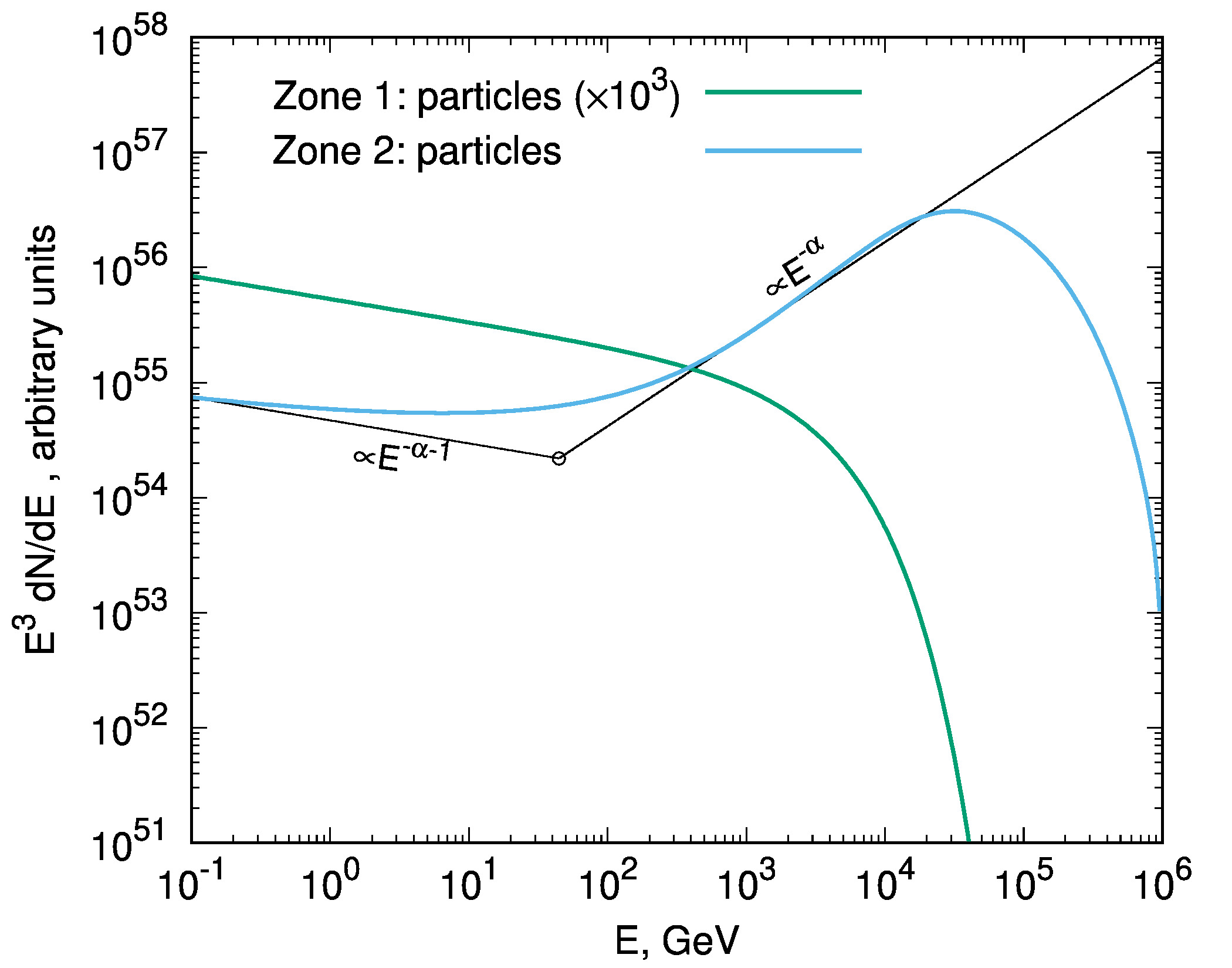}
\caption{Top panel: Synchrotron, IC cooling time together with the acceleration time.  Bottom panel: Electron distribution in two zones. Black guide lines indicate power-law approximations.\label{fig:cooling}}
\end{figure}

For the model calculations, magnetic field values of \(B_1=\bone\unit{G}\) and \(B_2=\btwo\unit{G}\) are assumed. The injection power is set to \(\sim\myL\unit{erg\,s^{-1}}\), and for the size of the production region we consider a value close to \(\rsize\unit{cm}\). If one considers this size in the context of a GRB afterglow, one should compare it to the forward shock radius, which depends on the time passed since the trigger, \(t\mysub{tr}'\):
\be
R\sim \Gamma^2t\mysub{tr}'c\sim 3\times10^{16}\qty(\frac{\Gamma}{10})^2\frac{t\mysub{tr}'}{3\unit{h}}\unit{cm}\,.
\ee
The typical energy density of the target photons in the production region is
\be
w\mysub{ph}\sim4\times10^{-5}\kappa_1\eta\mysub{rad}\qty(\frac{R}{3\times10^{16}\unit{cm}})^{-2}\unit{erg\,cm^{-3}}\,,
\ee
where \(\eta\mysub{rad}\) is the radiation efficiency in zone 1 (in what follows we ignore this factor, setting \(\eta\mysub{rad}=\myetar\), for the sake of simplicity). This energy density corresponds to an equivalent magnetic field strength of
\be
B\mysub{eq}\sim3\times10^{-2}\kappa_1^{\nicefrac{1}{2}}\qty(\frac{R}{3\times10^{16}\unit{cm}})^{-1}\unit{G}\,.
\ee
This photon field is the dominant target in zone 2, whereas it is negligible in zone 1. The corresponding cooling time scales are shown in Fig.~\ref{fig:cooling} (top panel). Whilst at high energies (approaching \(1\unit{PeV}\)), the Klein-Nishina losses  approach their asymptotic energy-dependence, \(\tau\mysub{kn}\propto E\), for the parameter set considered, the particles cool in the transition regime with \(\tau\propto\mathrm{const}\). Thus the spectrum formed is not as hard as expected from our earlier qualitative analysis.

The effect of the onset of Klein Nishina cooling on the electron spectrum is shown in Fig.~\ref{fig:cooling} (bottom panel), where the energy distribution of electrons in both zones are shown. For the calculations here we adopted the following parameters: linear size \(R=\rsize\unit{cm}\); total power of acceleration of non-thermal particles \(L_0=\myL\ergs\), which is distributed between the zones with \(\kappa_1=\kappaone\) and \(\kappa_2=\kappatwo\); the injection index \(\alpha=\myalpha\) (the ``main case''). Finally,  the acceleration efficiency was set to \(\eta\mysub{acc}=\myeta\), for which the cutoff energy in zone 1 is determined to be:
\be
E\mysub{cut}{}_{,1}\approx 6~\left(\frac{\eta\mysub{acc}}{10^2}\right)^{\nicefrac{-1}{2}}\qty(\frac{B_1}{1\unit{G}})^{\nicefrac{-1}{2}}\unit{TeV}\,.
\ee
For this acceleration efficiency the cutoff energy in zone 2 is at \(\approx200\unit{TeV}\), which is close to the energy at which the synchrotron losses dominate over the IC losses, \(E_*\approx20\unit{TeV}\), thus the influence of the high energy cutoff becomes prominent at energies just above the Klein-Nishina hardening energy scale. 

\begin{figure}
  \plotone{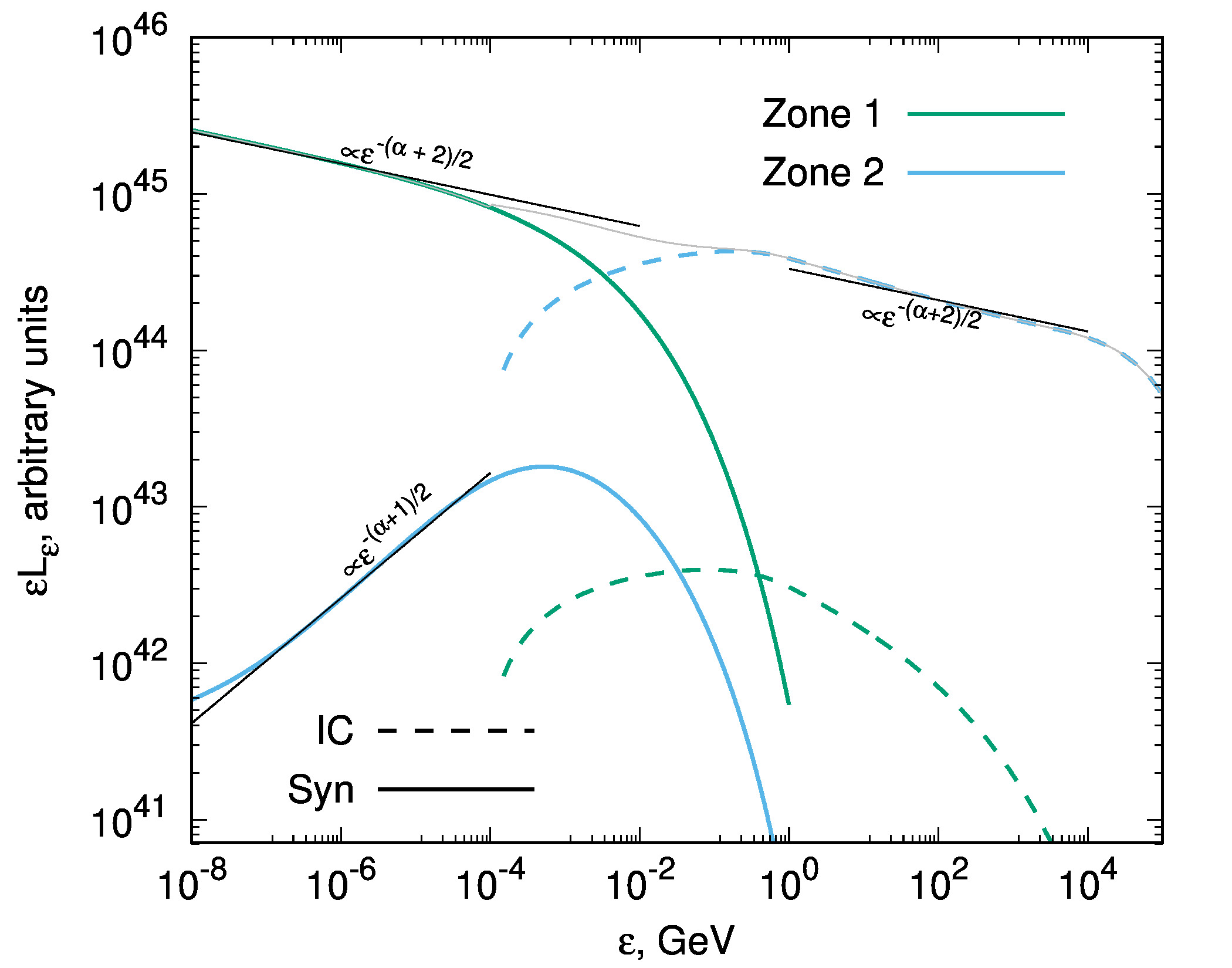}
  \caption{Spectral energy distribution of synchrotron and IC emission from two zones. Black guide lines show the power-law approximations.\label{fig:SED}}
\end{figure}

The energy dependence of the electron distribution is directly reflected in the synchrotron spectrum from zone 2. As can be seen from Fig.~\ref{fig:SED}, this component is subdominant to the luminous synchrotron component from zone 1. The photon index of the hardest part of the spectrum is \((\alpha+1)/2\approx1.5\), which is considerably softer than the limiting photon index of \(\gamma\mysub{s,kn}(=\alpha/2)\). 
This is caused by the smooth broad transition to the Klein-Nishina regime. While the broad transition from the Thomson to Klein-Nishina regimes causes the electron distribution to be not as hard as naively expected, the IC component appears to be somewhat harder than in the limiting case. As can be seen in Fig.~\ref{fig:SED}, a power-law component extends from a few \(\unit{GeV}\) to beyond \(10\unit{TeV}\) with a photon index of \(\approx(\alpha+1)/2\). Note that for our calculations we set \(\alpha=\myalpha\), and the production region bulk Lorentz factor was assumed to be \(\Gamma=\mygamma\). 

\begin{figure}[ht!]
\plotone{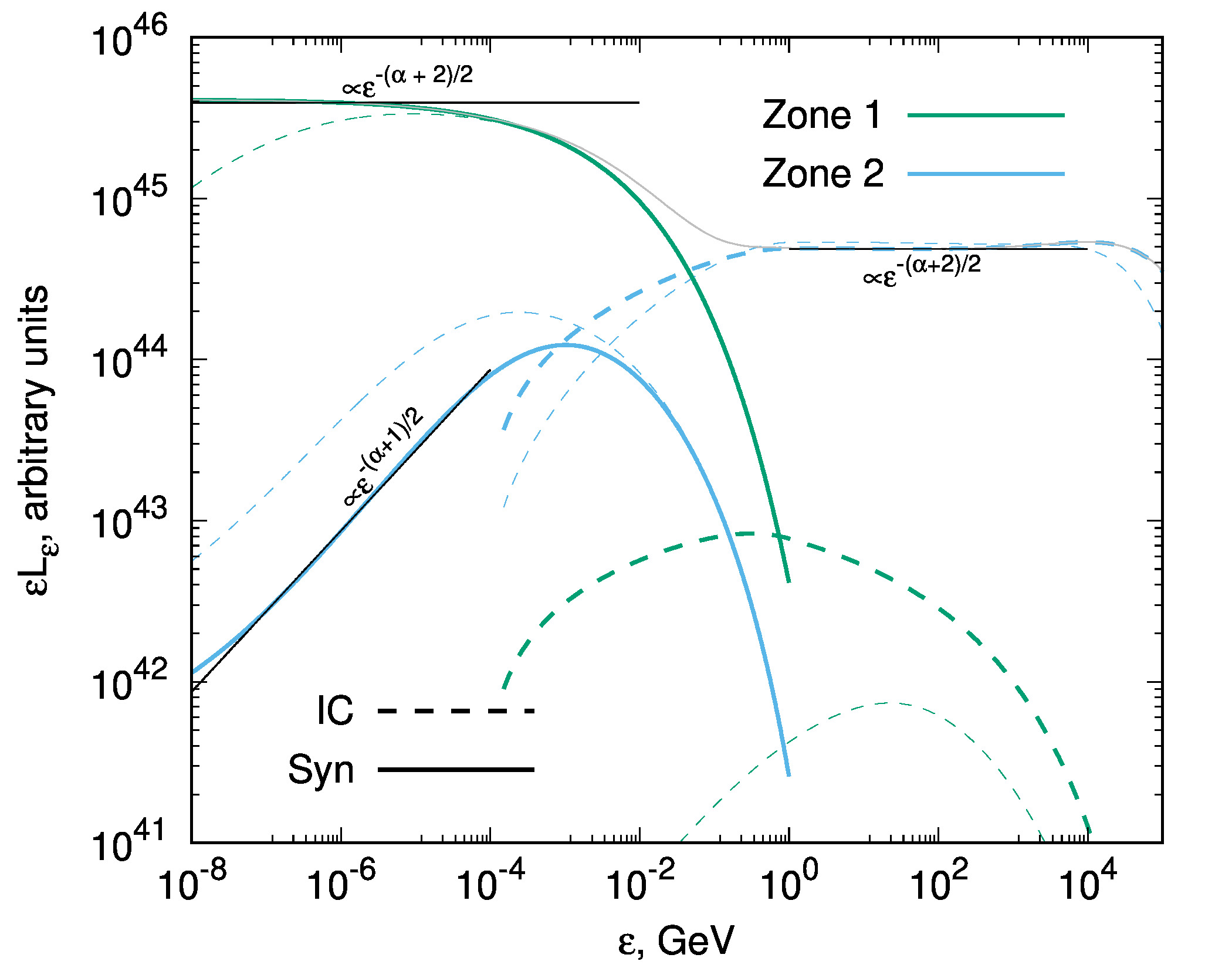}
\plotone{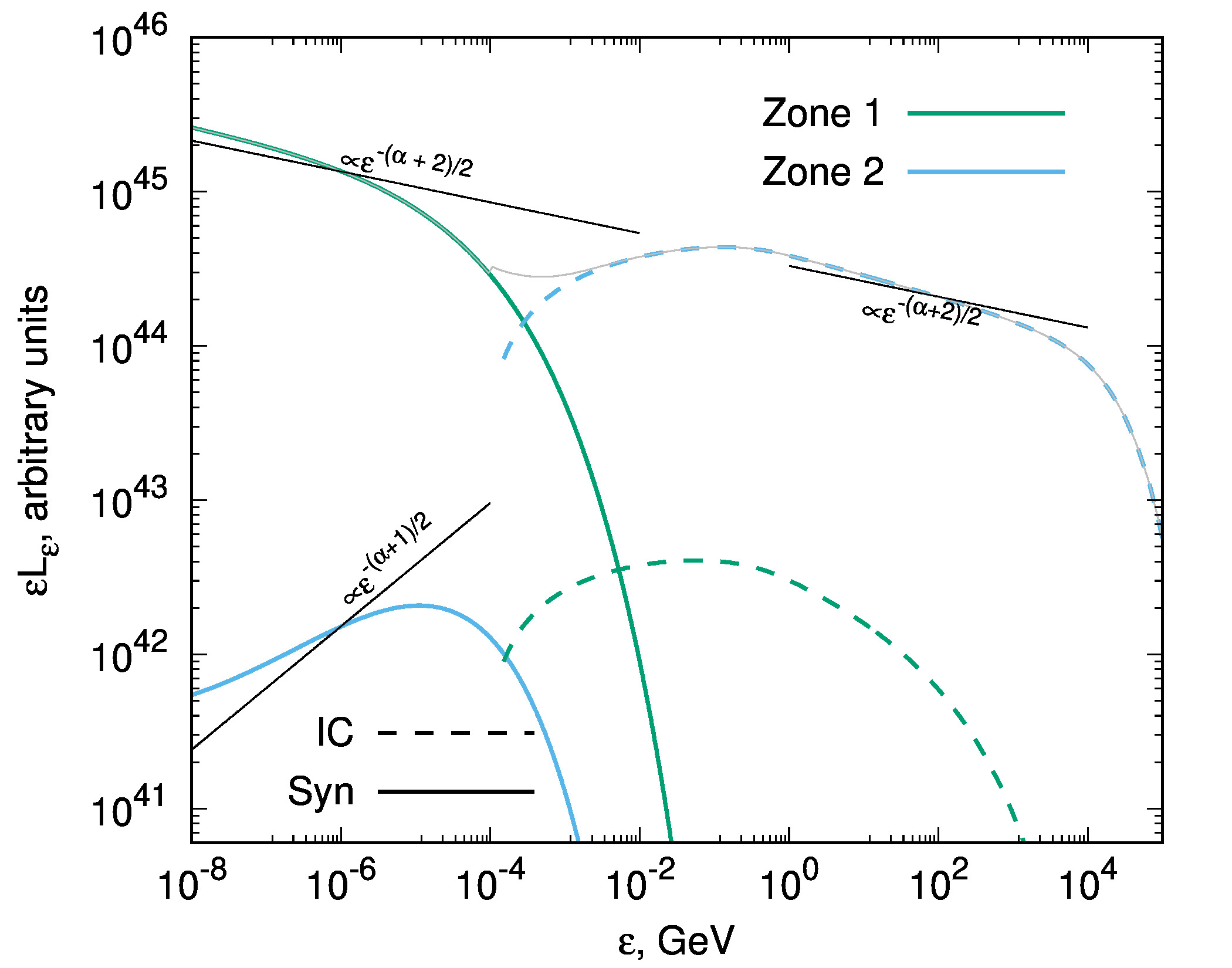}
\caption{Spectral energy distribution of synchrotron and IC emission from two zones. Black guide lines show the power-law approximations. Top panel: Case A; Bottom panel: Case B. Thin lines in the top panel correspond to a case when the electron distribution in the first zone features a cooling break at \(E\approx10\unit{GeV}\).\label{fig:SED_ab}}
\end{figure}

To illustrate the influence of the model parameters, we performed calculations for a range of different parameter sets. The results of these calculations are shown in Fig.~\ref{fig:SED_ab}. For the ``case A'' we adopted a different value for the injection index: \(\alpha=2\) instead of \(\alpha=2.2\) used in the ``main case''. For the ``case B''  we adopted a different value for the acceleration efficiency: \(\eta\mysub{acc}=10^4\) instead of \(\eta\mysub{acc}=10^2\) used in the ``main case''. The adopted model parameter values are summarized in Table~\ref{table:parameters}.

Low-energy target photons can be an important role in the formation of a hard VHE spectrum in the case of the conventional one-zone SSC models. To demonstrate the relatively small influence of low-energy target photons in the framework of our considered two-zone approach, in the top panel of Fig.~\ref{fig:SED_ab} we also plot the SED obtained under the same conditions assuming that the particle spectrum in the first zone features a cooling break at \(E\approx10\unit{GeV}\). The corresponding spectra are shown with thin lines. Under this assumption, the synchrotron spectrum from the first zone features the cooling breaks, as expected. The IC spectrum is more strongly suppressed: one sees here the impact of both the cooling break and reduced target photon density.  The reduction of the IC loss rate leads to a considerable enhancement of the synchrotron emission from the second zone (note that this component still remains subdominant). The IC spectra from the second zone shows, however, only minor changes, noticeably only close to the high- and low-energy cutoffs regions. This quite weak influence of the target photon spectrum on the spectral properties of the IC component from the second zone is caused by the fact that the IC losses determine the particle spectrum, as we assume that the emission is generated in the fast cooling regime. Therefore, the electron spectrum adjusts to the rate of the dominant losses, and the spectral properties of the IC component are largely determined by the injection spectrum.

\section{Discussion and Conclusion} \label{sec:discussion}

The need for studying energy losses in the inhomogeneous emission region downstream can be easily realized by considering the evolution of the magnetic field from the upstream to downstream regions. Based on the hydrodynamics of the forward shock propagating through the CBM, one can obtain the following estimate for the downstream magnetic field strength:
\be\label{eq:b_up}
B\sim 3\times10^2\frac{\Gamma}{10}\frac{B\mysub{cbm}}{10\unit{\mu G}}\unit{\mu G}\,.
\ee
This estimate depends on the typical strength of the CBM magnetic field, \(B\mysub{cbm}\), and accounts for the transformation of this field to the forward shock rest frame, and for the increase of the field strength at a weakly magnetized relativistic shock due to the shock compression.

The magnetic field given by Eq.~(\ref{eq:b_up}) appears significantly below the Gauss-level required for the afterglow radiation. Therefore, one needs to assume an efficient  magnetic field amplification process, which can increase the energy density of the magnetic field to the level comparable to the plasma energy density in the downstream:
\be
w\sim n\mysub{cbm}m_p\Gamma^2\approx0.15\qty(\frac{n\mysub{cbm}}{1\unit{cm^3}})\qty(\frac{\Gamma}{10})^2\unit{erg\,cm^{-3}}\,,
\ee
where \(n\mysub{cmb}\) is CBM density. This estimates shows that the magnetic field in the downstream can be amplified up to a strength of
\be
B\mysub{eq}=\sqrt{8\pi w}\sim 2\qty(\frac{n\mysub{cbm}}{1\unit{cm^3}})^{\nicefrac{1}{2}}\qty(\frac{\Gamma}{10})\unit{G}\,.
\ee
Gauss-strength magnetic fields in the afterglow production region are also favored on theoretical grounds by afterglow emission modeling. If the magnetic field is indeed amplified by a factor of \(\sim10^3\), it is natural to further assume that this amplification is inhomogeneous throughout the volume resulting in a magnetic field configuration with strong spatial fluctuations.
For example, magnetic field amplification by turbulent dynamo shows that the magnetic energy is predominantly localized in small blobs  \citep{2009ApJ...692L..40Z}. Moreover, this may be a general effect: the field amplification predominately operates on small scale fields \citep{1968JETP...26.1031K}.

The highly inhomogeneous structure of the downstream region can have important implications for the properties of the non-thermal emission generated. In particular, such a structure in the production region can significantly alter the synchrotron radiation emission, with clumps of highly amplified magnetic field leading to the synchrotron emission extending significantly beyond the one-zone synchrotron burn-off limit \citep{2021ApJ...914...76K}. This scenario requires that particles are accelerated in a region of weak magnetic field, and subsequently penetrate into a second zone of amplified magnetic field, where they rapidly cool producing VHE synchrotron radiation. The requirement of effective particle exchange between the two zones of strong and weak magnetic field is an important element of this scenario. 

It should be noted, however, that efficient particle exchange between the zones is a significant assumption. Processes exist, which can hinder particle exchange between the two zones. For example, if the change of the magnetic field strength is relatively smooth, the magnetic adiabatic invariant prevents particles from the zone of weak magnetic field reaching a strong magnetic field zone \citep[see the discussion in][]{2021ApJ...914...76K}. The particle escape from the zone of strong to weak magnetic field is not forbidden by the magnetic adiabatic invariant, but it seems feasible that one can neglect this process. Because of the much higher rate of synchrotron losses in the strong magnetic field zone, the total number of particles in this zone is naturally significantly reduced to that in the weak magnetic field zone, particularly for the highest energy particles with energies close to the maximum energy.
\begin{figure}
  \plotone{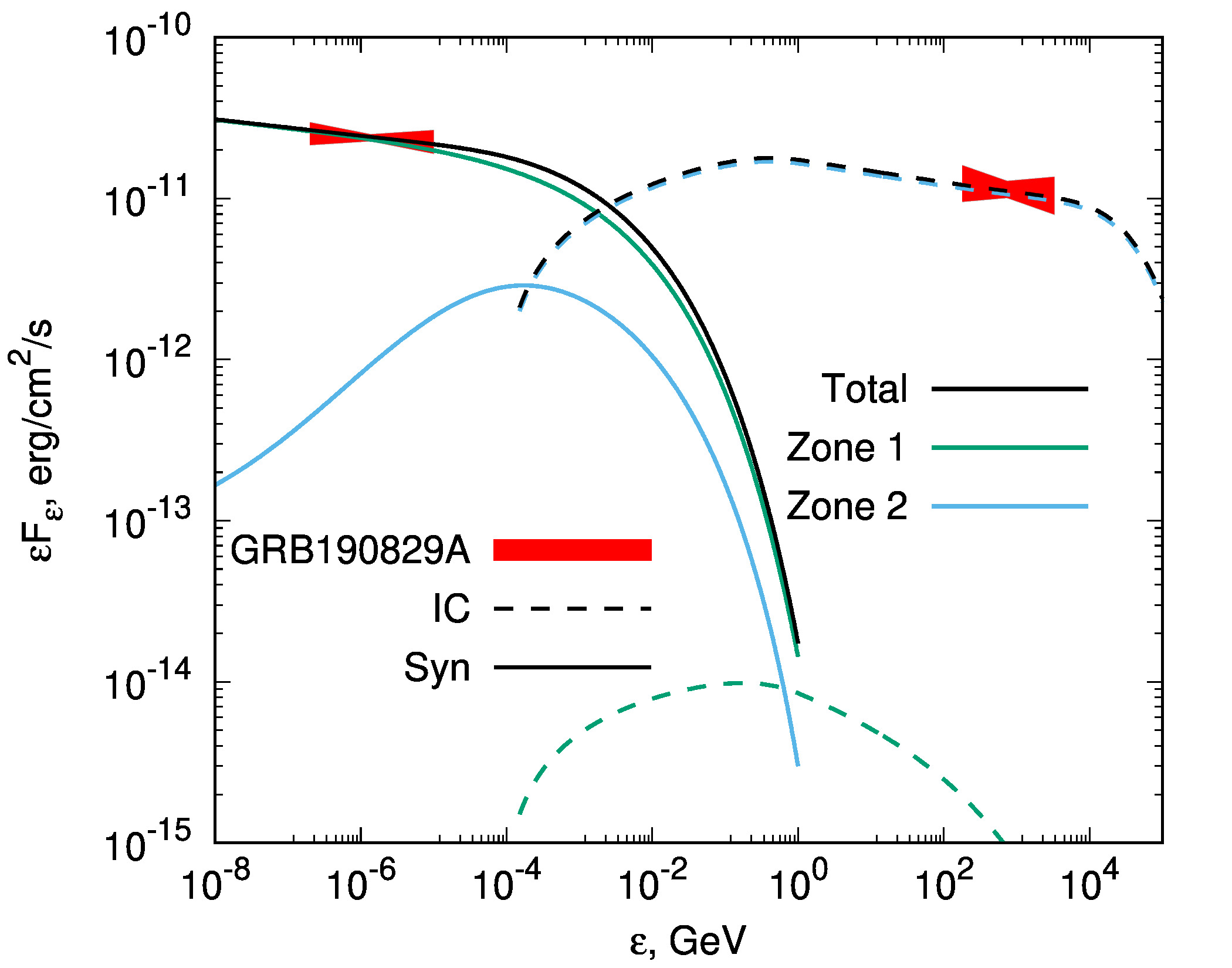}
  \caption{Spectral energy distribution of synchrotron and IC emission from two zones. Filled zones show X-ray (XRT) and VHE intrinsic (\hess) spectra \citep[for detail see in][]{2021Sci...372.1081H}.\label{fig:SED190829A}}
\end{figure}

On the other hand, synchrotron photons can freely travel between the two zones. The photon exchange between the zones have two major effects: (i) altering particle energy losses, and (ii) change the properties of IC emission. We suggest a simple model that allows one to study these two effects. We find that for feasible model parameters, IC scattering dominates the cooling process in the zone of weak magnetic field. Due to the Klein-Nishina effect, the particle spectrum formed in the fast cooling regime appears to be significantly harder than the spectrum formed for the case when synchrotron losses dominate. While the synchrotron emission from this zone may appear completely sub-luminous  with respect to the synchrotron emission generated in the strong magnetic field zone, the IC component from the weak magnetic field zone would be expected to dominate.

The second signature of the hard particle spectrum is expected in the IC component generated by these particles. This spectrum appears to be hard, with a photon index coinciding with the value expected for the synchrotron/Thomson spectra, \((\alpha+2)/2\) (where \(\alpha\) is the injection index). The IC spectrum therefore appears to have the same slope as the dominant synchrotron emission. The relative flux through these two channels is determined by the phenomenological parameters, \(\kappa_i\), which determine the ratio of the acceleration powers in the two zones. Our simulations  presented in Fig.~\ref{fig:SED190829A} show that for an acceleration spectrum with a spectral slope of \(\alpha=2.1\), which allows the slope of the X-ray spectra for \grbHII to be reproduced, the X-ray and VHE flux ratio seen in \grbHII implies \(\kappa_1=0.7\) and \(\kappa_2=0.3\)  (see column ``\grbHII'' in Table~\ref{table:parameters}). This suggests that acceleration processes of comparable power operate in the both zones. However, the acceleration in the zone of stronger magnetic field is somewhat more efficient. The obtained hard VHE IC spectrum extends beyond \(10\unit{TeV}\) for a modest bulk Lorentz factor of \(\Gamma=10\). This implies that a hard multi-TeV IC component can be generated also during the late afterglow phases, when the forward shock transits into the mildly relativistic regime. During the prompt or early afterglow phases, when the bulk Lorentz factor can be significantly larger, \(\Gamma\geq 100\), the intrinsically hard IC component can extend up to the ultra high energy domain (\(\geq100\unit{TeV}\)). However, we note that the extragalactic EBL attenuation is severe already in the VHE domain for even the most local GRB redshift values.

\begin{deluxetable*}{lcl}
\tablenum{1}
\tablecaption{Used notations\label{table:notation}}
\tablewidth{0pt}
\tablehead{\colhead{Parameter} & \colhead{Notation} & \colhead{Comment}
}
% \decimalcolnumbers
\startdata
Production region size&\(R\)& Similar to the forward shock radius\\
%Averaged magnetic field &\(\bar{B}\)&--\\
Magnetic field &\(B\)&\(B_{1,2}\) in zone 1 or 2; \(B\mysub{cbm}\) in the shock upstream\\ %; \(\bar{B}\) averaged\\
Bulk Lorentz factor &\(\Gamma\) & We do not distinguish the bulk Lorentz factors of emitting plasma or forward shock\\
Injection power &\(L_0\) & Eq.~(\ref{eq:normalization}) is in the co-moving frame, but note that  \(L_0\) is a Lorentz invariant\\
Normalization factors &\(\kappa_{1,2}\) & Power distribution between two zone; \(\kappa_1+\kappa_2=1\)\\
Energy density &\(w\)& e.g., \(w\mysub{ext}\) is energy density of external photon fields\\
Electron energy & \(E\) & --\\
Electron energy density & \(n\) & i.e., \(\dd{N}=n\dd{E}\)\\
Injection index &\(\alpha\) &see, e.g., Eq.~\ref{eq:injection_spectrum}\\
Injection rate & \(q\) & i.e., \(\dd{N}=q\dd{E}\dd{t}\)\\
Acceleration efficiency&\(\eta\mysub{acc}\)&see, e.g., Eq.~(\ref{eq:max_energy})\\
Cooling time &\(\tau\)& Synchrotron, IC, and adiabatic cooling or escape\\
Energy loss rate &\(\dot{E}\)& Synchrotron, IC, and adiabatic\\
Photon energy & \(\ve\) & For photons produced through the synchrotron or IC channels\\
Photon index & \(\gamma\) & In particular, \(\gamma\mysub{s}\), \(\gamma\mysub{kn}\)\\
Radiation  efficiency&\(\eta\mysub{rad}\)& We assume \(\eta\mysub{rad}=1\)\\
Circumburst medium density&\(n\mysub{cmb}\)& i.e. \(\dd{N}\mysub{cmb}=n\mysub{cmb}\dd{V}\)\\
\enddata
%\tablecomments{}
\end{deluxetable*}

\begin{deluxetable*}{lcccccc}
\tablenum{2}
\tablecaption{Parameter values used for model calculations\label{table:parameters}}
\tablewidth{0pt}
\tablehead{\colhead{Parameter} & \colhead{Notation} & \colhead{units} &\colhead{Main case} &\colhead{Case A} &\colhead{Case B} &\colhead{\grbHII}
}
% \decimalcolnumbers
\startdata
 & & &in Figs.~\ref{fig:cooling}\&\ref{fig:SED}&in Fig.~\ref{fig:SED_ab}&in Fig.~\ref{fig:SED_ab}&in Fig.~\ref{fig:SED190829A}\\
Size & \(R\)& \unit{cm} & \(10^{16}\)& \(10^{16}\) & \(10^{16}\) & \(10^{16}\) \\
Bulk Lorentz factor&\(\Gamma\)&--&\(10\)&\(10\)&\(10\)&\(10\)\\
Luminosity&\(L_0\)&\ergs&\(10^{39}\)&\(10^{39}\)&\(10^{39}\)&\(2.2\times10^{38}\)\\
Strong magnetic field&\(B_1\)&\unit{G}&\(1\)&\(1\)&\(1\)&\(1\)\\
Weak magnetic field&\(B_2\)&\unit{G}&\(10^{-3}\)&\(10^{-3}\)&\(10^{-3}\)&\(10^{-3}\)\\
Zone 1 power fraction&\(\kappa_1\)&--&\(0.9\)&\(0.9\)&\(0.9\)&\(0.7\)\\
Zone 2 power fraction&\(\kappa_2\)&--&\(0.1\)&\(0.1\)&\(0.1\)&\(0.3\)\\
Injection slope&\(\alpha\)&--&2.2&2&2.2&2.1\\
Acceleration efficiency&\(\eta\mysub{acc}\)&--&\(10^2\)&\(10^2\)&\(10^4\)&\(10^2\)\\
Radiation efficiency&\(\eta\mysub{rad}\)&--&\(1\)&\(1\)&\(1\)&\(1\)\\
Distance&\(D\)&\unit{Mpc}&--&--&--&\(400\)\\
\enddata
%\tablecomments{}
\end{deluxetable*}

\begin{acknowledgments}
DK acknowledges support by the RSF grant No. 21-12-00416.
\end{acknowledgments}

\vspace{5mm}

\bibstyle{aasjournal}



\end{document}